\documentclass[journal]{IEEEtran}

\usepackage{graphicx}

\begin{document}

\title{Comments on ``Compression of 3D Point Clouds Using a Region-Adaptive Hierarchical Transform''}

\author{Gustavo Sandri,
        Ricado~L. de~Queiroz,
        and~Philip~A. Chou
\thanks{G. Sandri is with Federal Institute of Brasilia, Brazil, \newline e-mail: gustavo.sandri@ifb.edu.br}
\thanks{R. de Queiroz is with the Department of Computer Science, University of Brasilia, Brazil, e-mail: queiroz@ieee.org}
\thanks{P. Chou is with 8i Labs Inc., Bellevue, Wa, USA, e-mail pachou@ieee.org}
\thanks{}}

\markboth{}%
{}
%



\maketitle

\begin{abstract}
The recently introduced coder based on region-adaptive hierarchical transform (RAHT) for the compression of point clouds attributes \cite{Queiroz:raht}, was shown to have a performance competitive with the state-of-the-art, while being much less complex. 
In \cite{Queiroz:raht}, top performance was achieved using arithmetic coding (AC), while adaptive run-length Golomb-Rice (RLGR) coding was presented as a lower-performance lower-complexity alternative. 
However, we have found that by reordering the RAHT coefficients we can largely increase the runs of zeros and significantly increase the performance of the RLGR-based RAHT coder. As a result, the new coder, using ordered coefficients, was shown to outperform all other coders, including AC-based RAHT, at an even lower computational cost.
We  present new results and plots that should enhance those in \cite{Queiroz:raht} to include the new results for RLGR-RAHT. 
We risk to say, based on the results herein, that RLGR-RAHT with sorted coefficients is the new state-of-the-art in point cloud compression.
\end{abstract}


%
\IEEEpeerreviewmaketitle

\section{Introduction}

\IEEEPARstart{T}{he} region-adaptive hierarchical transform (RAHT) was recently introduced \cite{Queiroz:raht} in order to compress color information and other attributes of point clouds. 
The RAHT is a hierarchical sub-band transform that resembles an adaptive variation of a Haar wavelet.
It uses an entropy coder to encode RAHT coefficients based on arithmetic coding (AC). 
The reported results demonstrate a performance which is comparable to that of the then state-of-the-art graph transform (GT) based coder \cite{Zhang:2014}, but at a much lower complexity. 
Adaptive run-length Golomb-Rice encoding (RLGR)~\cite{Malvar:RLGR}, which is a much less complex entropy coder, was also tested with RAHT. 
Results were shown \cite{Queiroz:raht} that RLGR-based RAHT (or RLGR-RAHT) was noticeably inferior to AC-based RAHT (AC-RAHT), although at a lower complexity.  

We want to show that by rearranging RAHT coefficients, one may improve the RLGR-RAHT performance to the point of outperforming not only RLGR-RAHT in \cite{Queiroz:raht} but also the AC-RAHT. 
With the RLGR-RAHT with sorted coefficients we will show that its performance improves over the state-of-the-art in point cloud compression at a reduced cost compared to AC-RAHT or GT-based coders \cite{Zhang:2014},\cite{klt}.

\section{RAHT coefficient ordering}

\begin{figure}[hbt]
	\center
	\begin{tabular}{c}
		\includegraphics[width=0.8\linewidth]{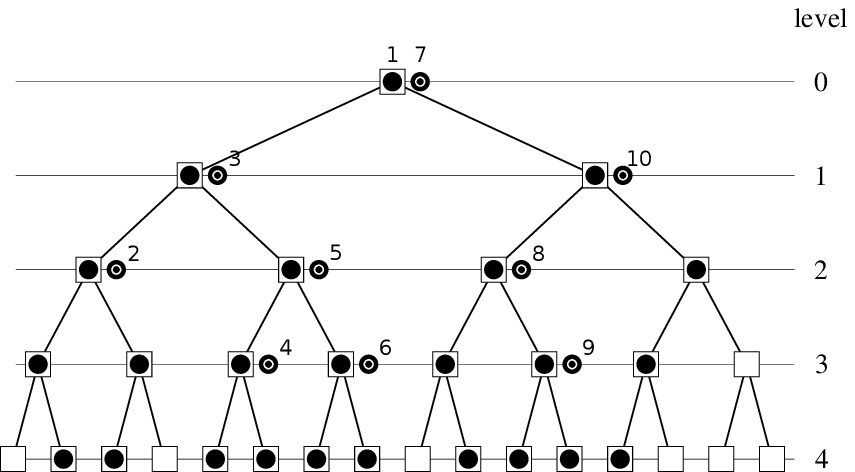} \\
		\includegraphics[width=0.8\linewidth]{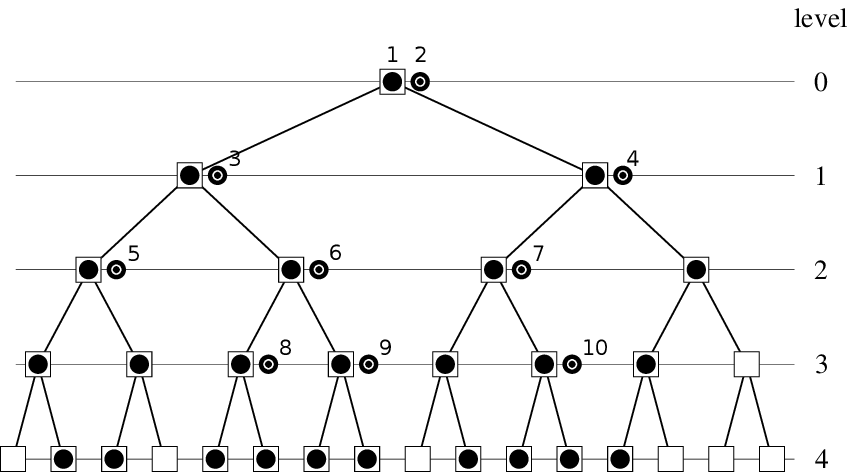} \\
		\includegraphics[width=0.8\linewidth]{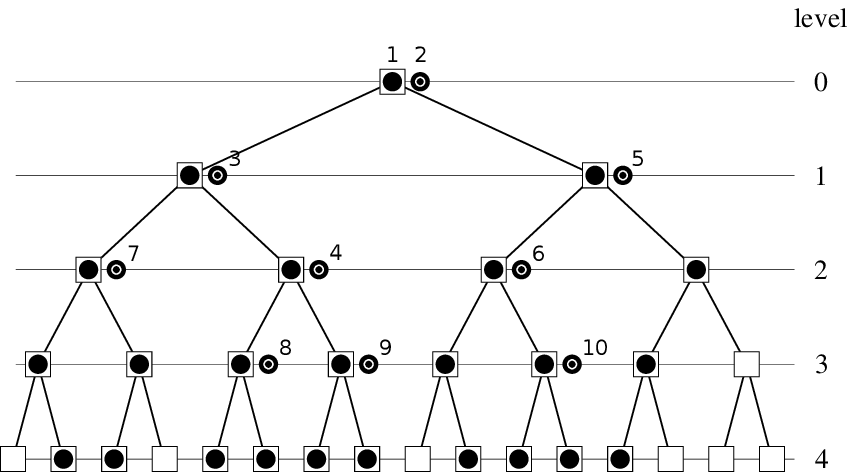}
	\end{tabular}
	\caption{RAHT decomposition and ordering illustration. After the transform, the order of DC and high-pass coefficients is indicated for the unsorted, depth- and weight-sorted cases. }
	\label{fg:ordering}
\end{figure}

RAHT is implemented by following backwards the octtree scan, from individual voxels to the entire voxel space, at each step recombining voxels into larger ones until reaching the root (see Fig.~\ref{fg:ordering}).
In Fig.~\ref{fg:ordering}, we start at level 4. 
If neighbor voxels in the same branch are occupied their colors are combined through a linear transformation (\cite{Queiroz:raht}, Eq.~(6)) and promoted to level 3. 
If a voxel does not have a neighbor in the same branch it is promoted straight to level 3. 
This process is repeated at each level until reaching level 0, the octtree root.
In the figure, once a pair of coefficients (or voxels) is transformed it generates a high-pass coefficient which is ready to be quantized and encoded, and a low-pass coefficient which is passed along to the next level of the tree. 
The decoder needs the DC coefficient (tree root) and all the high-pass coefficients generated during the transform.
There exists numerous ways to order the coefficients.
RAHT original implementation sends the DC component first, as shown in Fig.~\ref{fg:ordering}, and traverses the  leftmost branches until arriving at a leaf. 
It then repeats the scan towards the right side. 
Effectively it scans the high-passes from left to the right of the tree generating an unsorted list of coefficients. 
We can also scan the high-pass coefficients sorted by depth of the tree or sorted by weights (a weight of a coefficient represents how many vocels were involved in generating it \cite{Queiroz:raht}).

The linear transformation of the voxel pair produces a low- and a high-pass component, similar to the Haar transform. 
Hence, RAHT coefficients generated deeper in the tree (from root to leaves) represent higher frequencies.
For natural images, however, we expect the high frequency components to have lower amplitude than low frequency ones. 
Thus, quantizing high frequency coefficients may lead to a larger number of zero-valued quantized coefficients compared to low-frequency ones. 

We  propose to reorder the RAHT coefficients in ascending order according to their associated depth (depth of the octtree where they were generated) and to entropy encode the quantized and reordered coefficients using RLGR.
The idea of reordering the RAHT coefficients prior to entropy coder is not new and was also proposed in \cite{Pavez:2017}. 
In \cite{Pavez:2017}, RAHT coefficients were sorted by their weight in descending order and encoded using RLGR.

\section{Experimental Results}

\begin{figure}[hbt]
	\center
	\includegraphics[width=\linewidth]{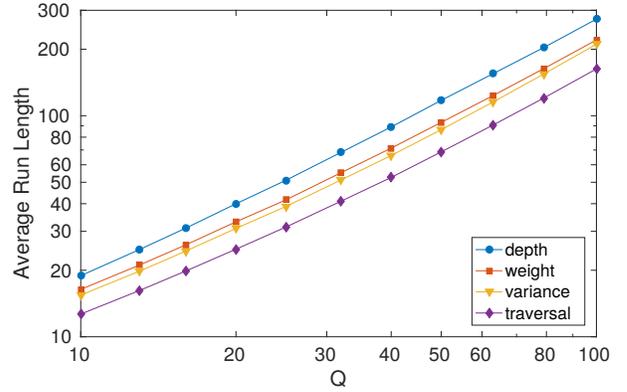}
	\caption{Average zero-run-length among the 7 test point clouds, using different sorting criteria.}
	\label{fg:arl}
\end{figure}

\begin{figure}[hbt]
	\center
	\begin{tabular}{c}
		\includegraphics[width=\linewidth]{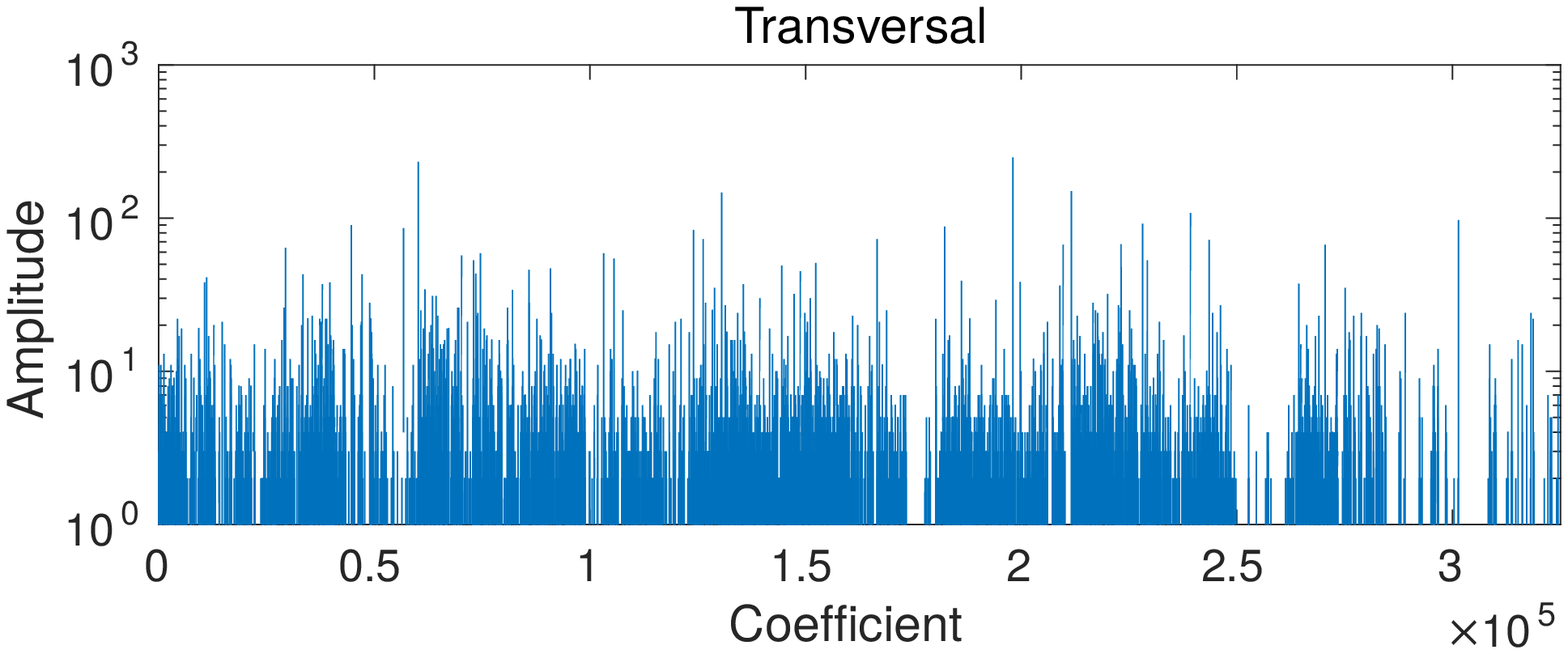} \\
		\includegraphics[width=\linewidth]{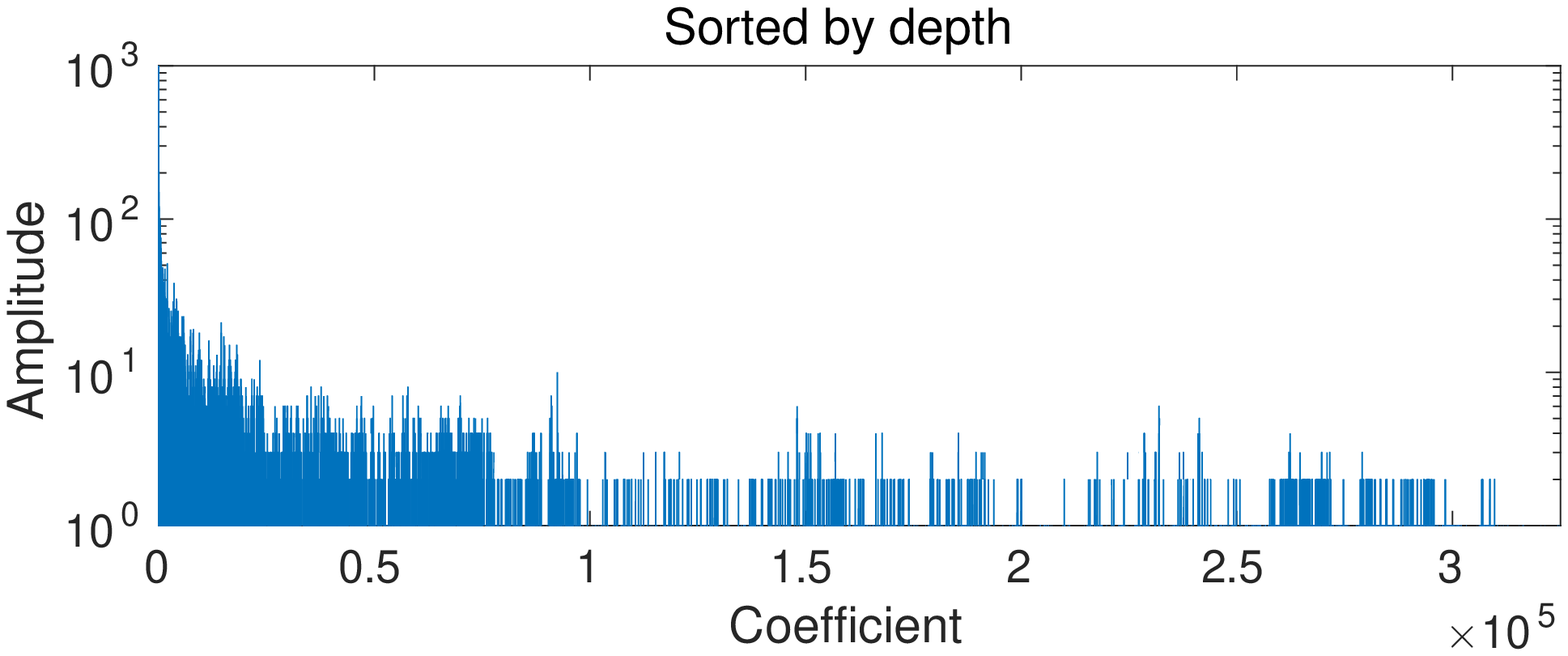}
	\end{tabular}
	\caption{Unsorted and sorted quantized coefficients amplitude for point cloud ``Phil'' ($Q = 20$). The unsorted coefficients present a random-like pattern, while the sorted ones have a somewhat decaying amplitude pattern.}
	\label{fg:coef}
\end{figure}

The proposed ordering induces the generation of longer zero sequences. 
Figure~\ref{fg:arl} compares the average zero-run-length for the $7$ test point clouds used in \cite{Queiroz:raht} when subject to RAHT and given quantizer step, for three different sorting methods: by weight, depth and unsorted. 
One can easily see that tree-depth ordering yields longer zero-run-lengths.  
Furthermore, RLGR is an adaptive algorithm that continuously updates its parameters. 
Figure~\ref{fg:coef} compares unsorted and sorted coefficients for frame ``Phil'', where we can expect that the better-behaved decaying pattern in the depth-sorted case may lead to faster adaptation and better compression with the RLGR algorithm.

Tables \ref{tb:RLGR10} and \ref{tb:RLGR40} compare sorting criteria in RLGR-RAHT. 
In them, and the rest of this work, rate is given in bits per occupied voxel (bpv) and distortion in peak signal-to-noise ratio (PSNR in dB) comparing the luminance component of the original and reconstructed frames.

\begin{table}[hbt]
	\center
	\caption{Rate comparison of the RLGR performance when using different sorting criteria ($Q = 10$).}
	\label{tb:RLGR10}
	\begin{tabular}{l|c c c c|c}
		\multicolumn{1}{c}{} & \multicolumn{4}{c}{\textbf{Bits per voxel}} \\
		\multicolumn{1}{c|}{\textbf{Point cloud}} & \textbf{transversal} & \textbf{depth} & \textbf{weight} & \textbf{variance} & \textbf{PSNR} \\
		\hline
		Man		& 3.1046 & 2.0581 & 2.1691  & 2.2145 & 40.3260\\
		Andrew	& 4.4661 & 3.3881 & 3.4893  & 3.5149 & 39.8398\\
		Phil	& 5.0985 & 3.7117 & 3.8018  & 3.8143 & 39.7470\\
		Ricardo	& 1.5040 & 0.8993 & 1.0137  & 1.0023 & 43.6176\\
		Sarah	& 1.9500 & 1.1003 & 1.2203  & 1.2422 & 43.4137\\
		Skier	& 4.7303 & 2.8717 & 3.0303  & 3.0432 & 40.7751\\
		Objects	& 4.2951 & 3.3085 & 3.4579  & 3.5029 & 39.9419\\
	\end{tabular}
\end{table}

\begin{table}[hbt]
	\center
	\caption{Rate comparison of the RLGR performance when using different sorting criteria ($Q = 40$).}
	\label{tb:RLGR40}
	\begin{tabular}{l|c c c c|c}
		\multicolumn{1}{c}{} & \multicolumn{4}{c}{\textbf{Bits per voxel}} \\
		\multicolumn{1}{c|}{\textbf{Point cloud}} & \textbf{transversal} & \textbf{depth} & \textbf{weight} & \textbf{variance} & \textbf{PSNR} \\
		\hline
		Man		& 0.7638 & 0.5345 & 0.5570 & 0.5577 & 31.5615\\
		Andrew	& 1.0225 & 0.7437 & 0.7715 & 0.7698 & 29.7742\\
		Phil	& 1.0416 & 0.6904 & 0.7082 & 0.7090 & 30.6536\\
		Ricardo	& 0.3395 & 0.2074 & 0.2254 & 0.2120 & 37.0116\\
		Sarah	& 0.4474 & 0.2455 & 0.2772 & 0.2784 & 36.3494\\
		Skier	& 1.3744 & 0.9071 & 0.9450 & 0.9366 & 31.6845\\
		Objects	& 0.8551 & 0.6692 & 0.7043 & 0.6882 & 31.8801\\
	\end{tabular}
\end{table}

\begin{figure}[hbt]
	\center
	\begin{tabular}{c}
		\includegraphics[width=0.945\linewidth]{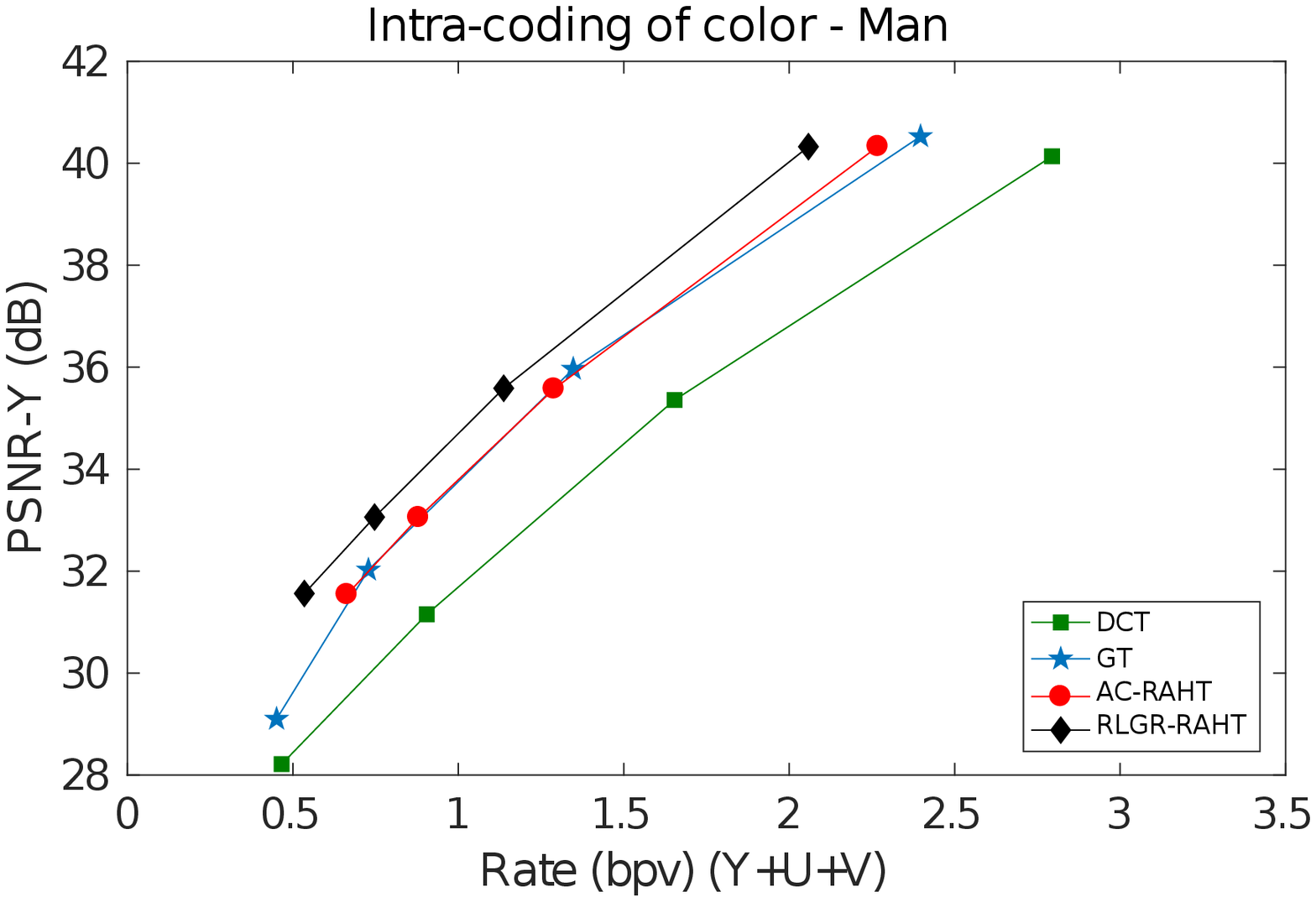} \\
		\includegraphics[width=0.945\linewidth]{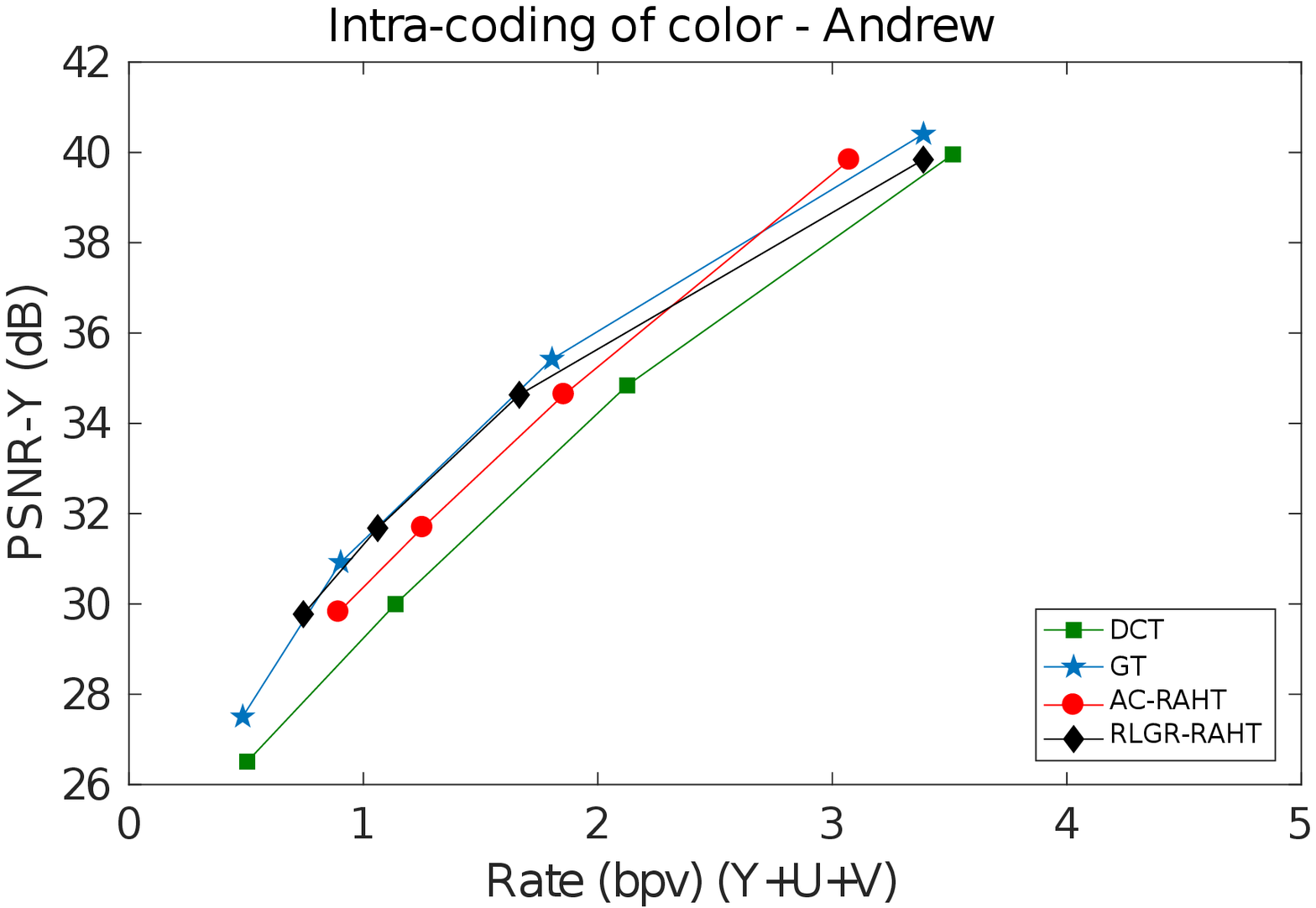} \\
		\includegraphics[width=0.945\linewidth]{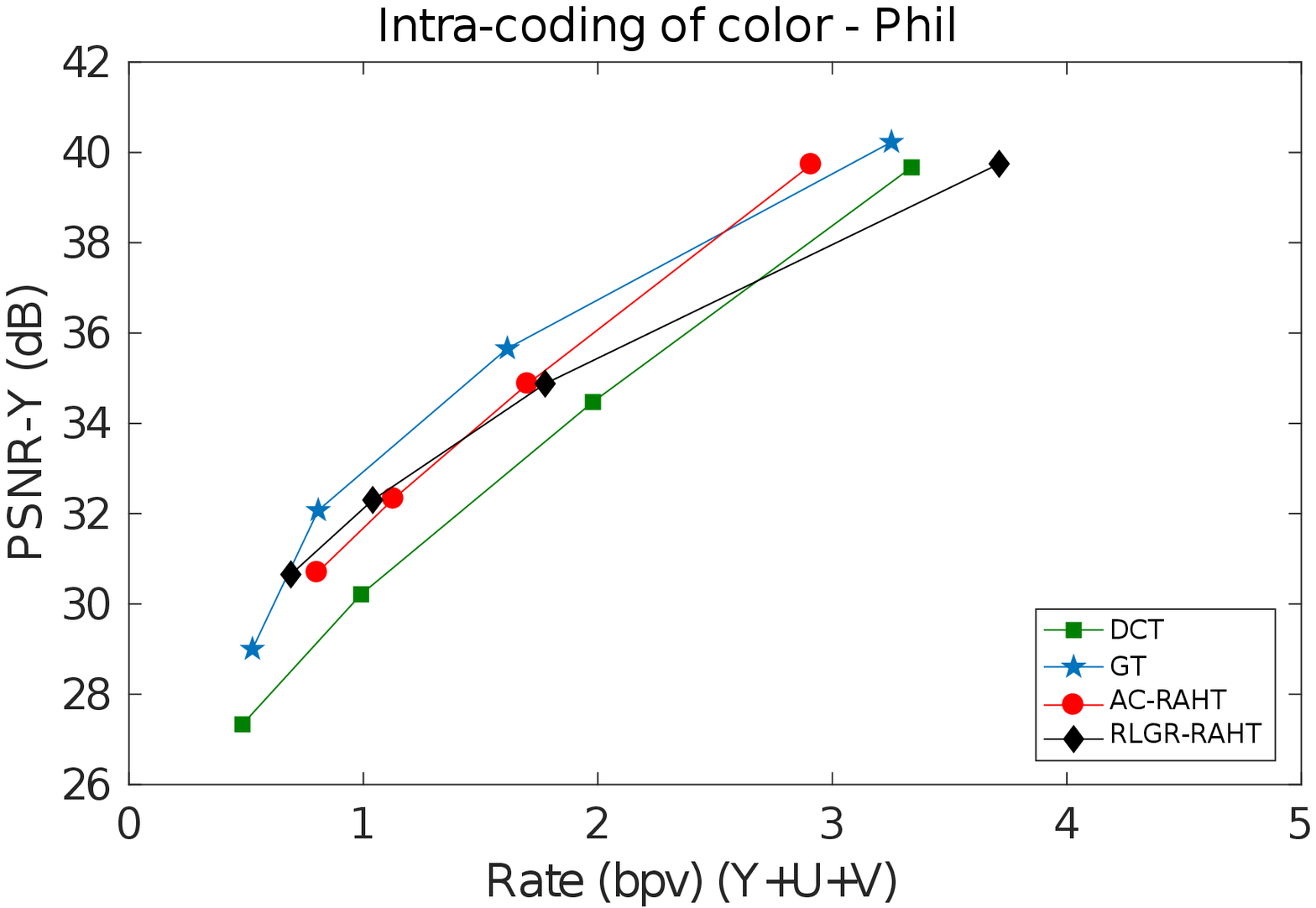}
	\end{tabular}	
	\caption{Rate-distortion curves, for different point clouds and coders.}
	\label{fg:rd_1}
\end{figure}

\begin{figure}[hbt]
	\center
	\begin{tabular}{c}
		\includegraphics[width=0.945\linewidth]{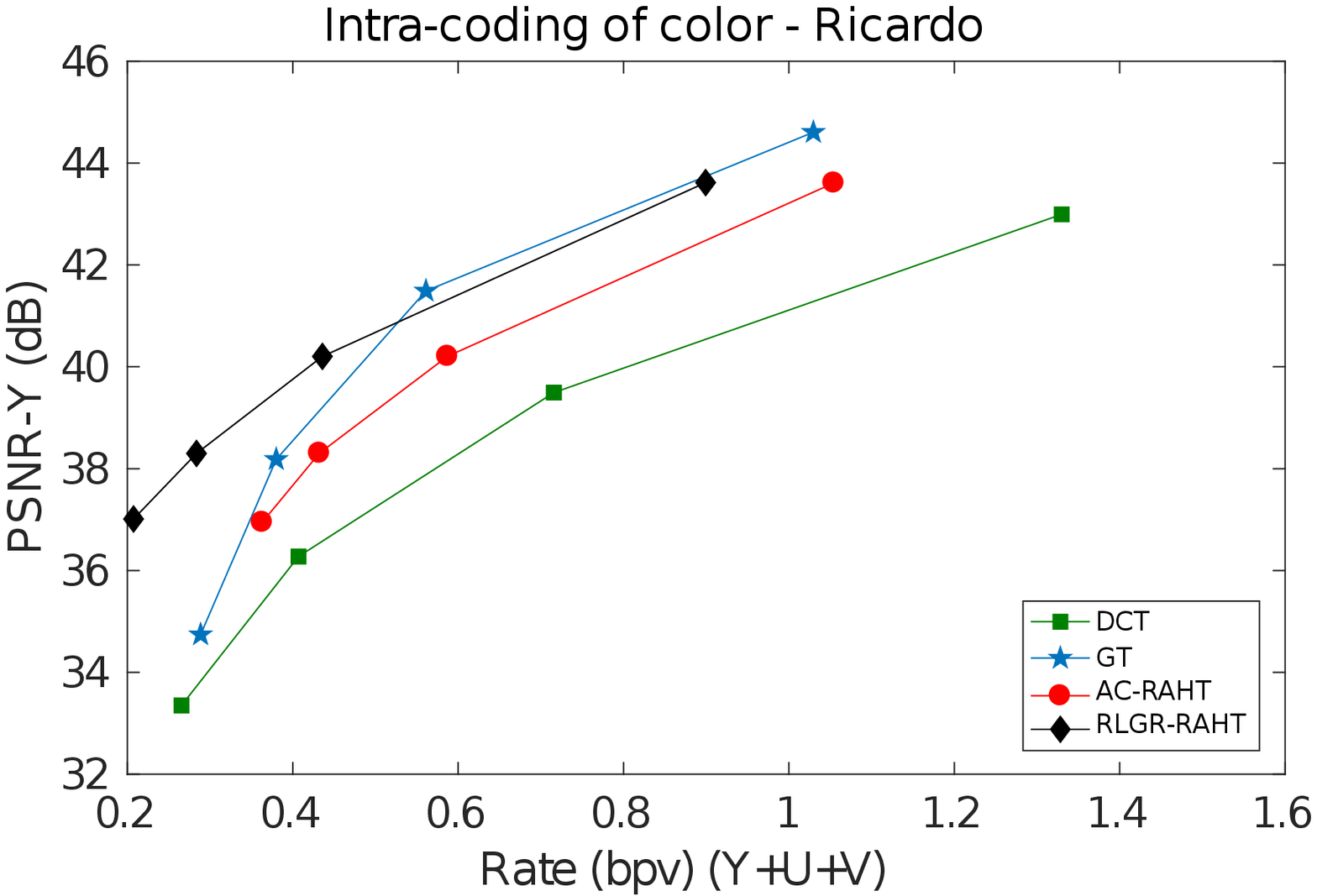} \\
		\includegraphics[width=0.945\linewidth]{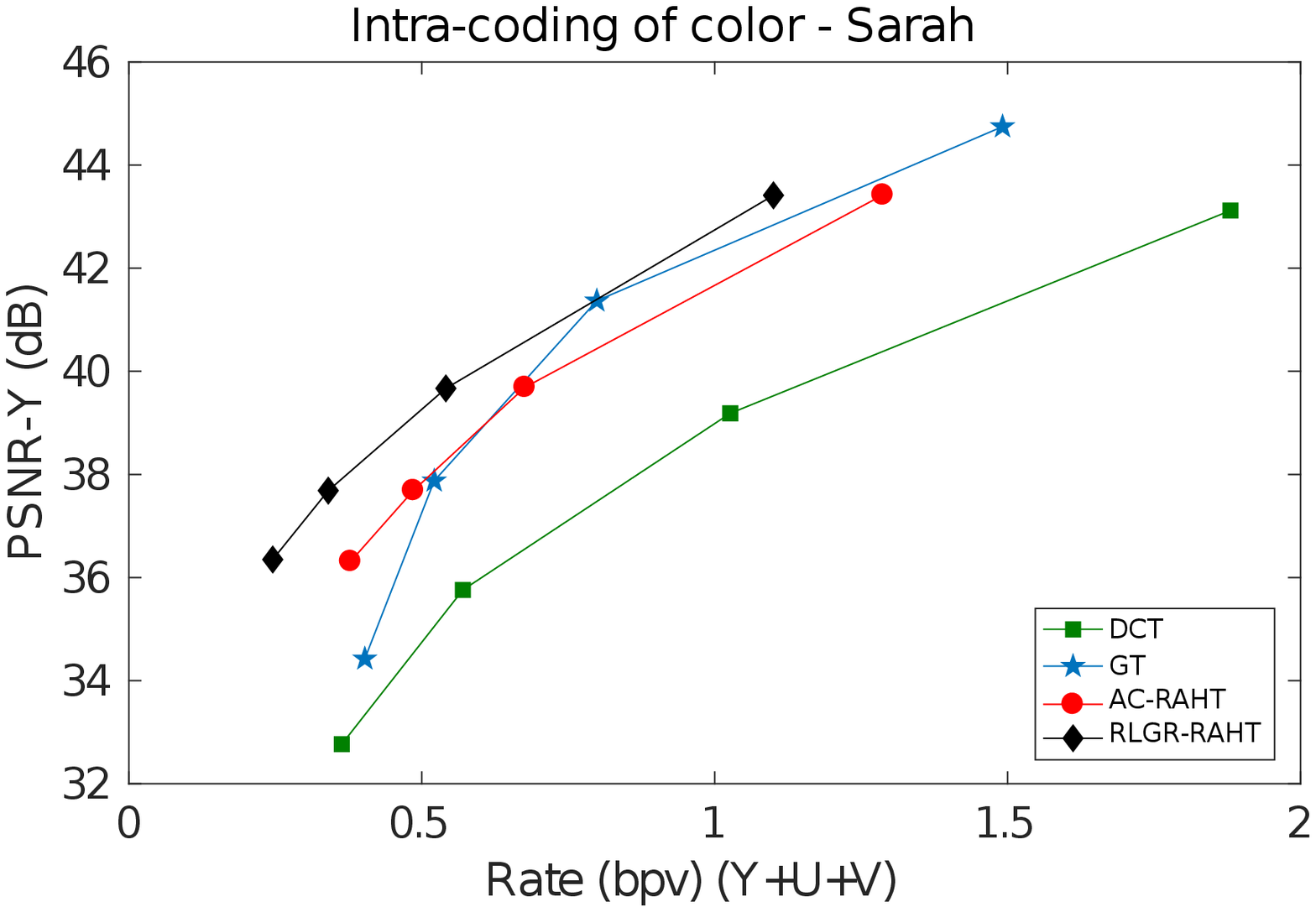} \\
		\includegraphics[width=0.945\linewidth]{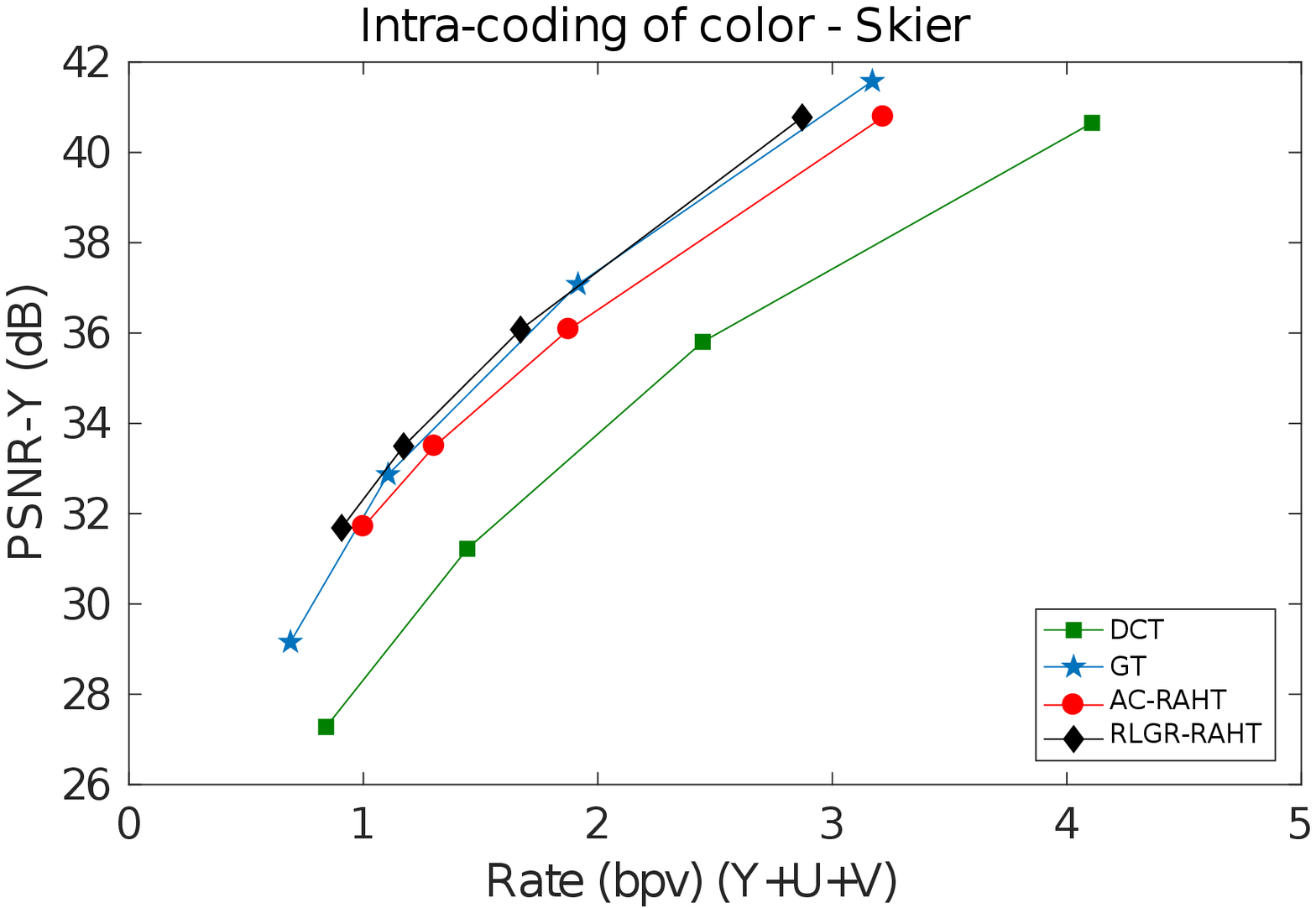} \\
		\includegraphics[width=0.945\linewidth]{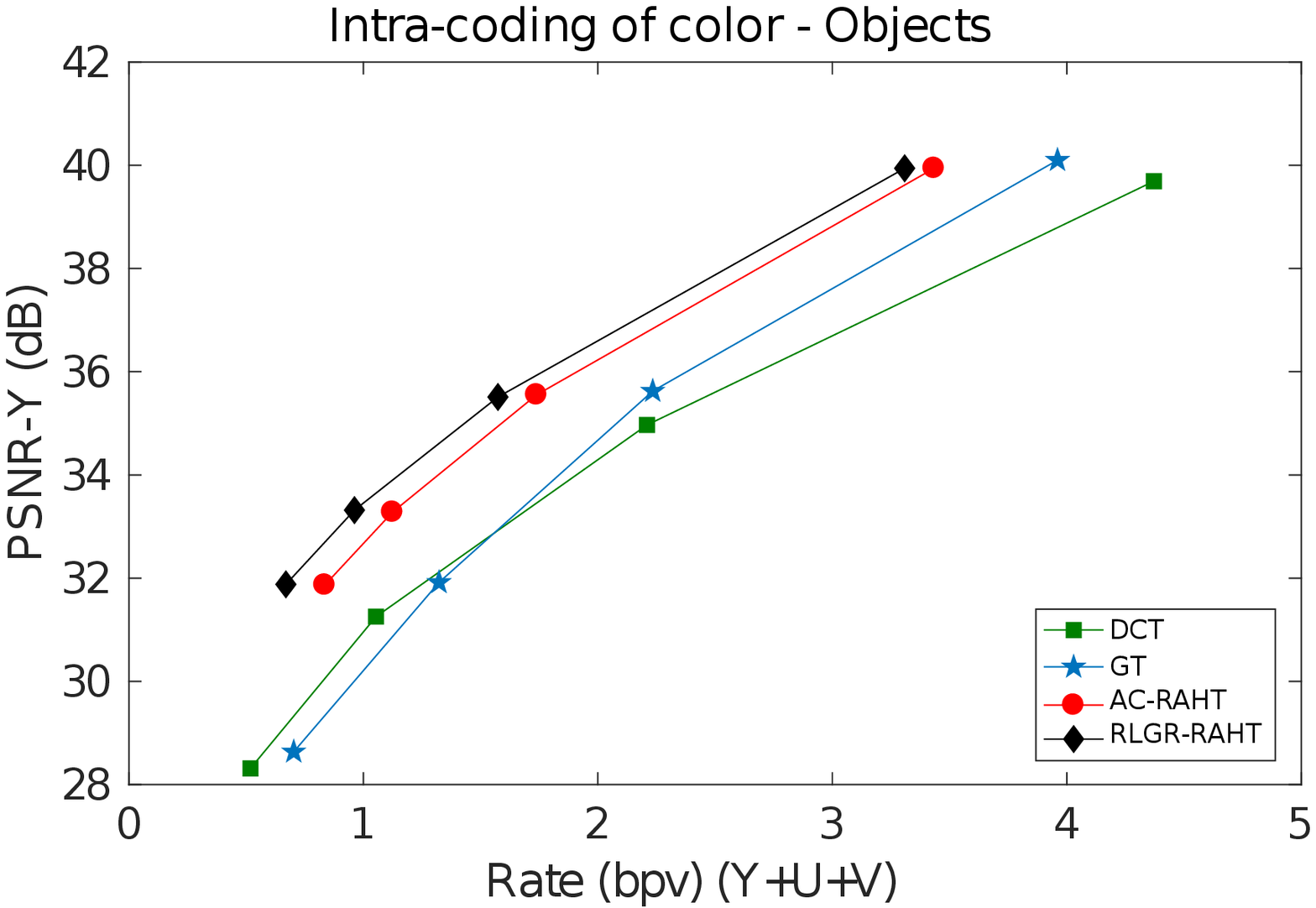}
	\end{tabular}	
	\caption{Rate-distortion curves, for different point clouds and coders.}
	\label{fg:rd_2}
\end{figure}

We can see from the tables that the proposed depth-ordered sorting of the RAHT coefficients yields better performance compared to weight-ordered or unordered methods. 
In light of this, we run the proposed depth-ordered RLGR-RAHT for all images and quantizer steps in the previous work and we believe that Figs. 6, 7 and 8(a) in \cite{Queiroz:raht} should be replaced with the improved results in Figs.~\ref{fg:rd_1} and \ref{fg:rd_2} in this work.

From the many rate-distortion curves, it is clear that depth-ordered RLGR-RAHT yields the best performance, even outperforming the GT-based coder. 

\section{Conclusion}

We have presented an ordering criteria to improve the performance of the RAHT entropy coder for 3D point cloud attributes. 
This transform has an almost linear complexity and can be easily computed in real time. 
The RLGR also has low complexity. 
The proposed method has a competitive performance, even outperforming the much more complex GT-based point cloud coder.
We believe the proposed depth-ordered RLGR-RAHT encoder presents the best performance so far in the compression of voxelized point clouds, with better rate-distortion trade-offs and lower complexity than rival approaches.

\end{document}